\documentclass[apj,twocolumn]{openjournal}
\usepackage{natbib} 
\usepackage[dvipsnames]{xcolor}
\usepackage{aas_macros} 
\usepackage{amssymb}
\usepackage{amsmath}
\usepackage[title]{appendix}
\usepackage{hyperref}	
\hypersetup{colorlinks=true,linkcolor=blue,citecolor=blue,filecolor=blue,urlcolor=blue}
\usepackage[caption=false]{subfig}
\usepackage{soul}                          
\usepackage{xspace}
\usepackage{booktabs}
\usepackage{savesym}
\savesymbol{tablenum}
\usepackage{siunitx}
\usepackage{pifont}
\restoresymbol{SIX}{tablenum}

\DeclareSIUnit\Msun{\ensuremath{M_\odot}}
\DeclareSIUnit\Zsun{\ensuremath{Z_\odot}}
\DeclareSIUnit\hred{\ensuremath{\textit{h}}}

\begin{document}
\title[]{MEGATRON: The Physical Origins of Steep UV Slopes at high redshift\vspace{-15mm}}
\author{Harley Katz$^{1,2*}$, Nicholas Choustikov$^{3}$, Aayush Saxena$^{3}$, Corentin Cadiou$^{4}$, Martin P. Rey$^{5}$, Jeremy Blaizot$^{6}$, Alex J. Cameron$^{7,8}$, Fergus Cullen$^{9}$, Taysun Kimm$^{10}$, \& Kosei Matsumoto$^{11}$}
\thanks{$^*$E-mail: \href{mailto:harleykatz@uchicago.edu}{harleykatz@uchicago.edu}}

\affiliation{$^{1}$Department of Astronomy \& Astrophysics, University of Chicago, 5640 S Ellis Avenue, Chicago, IL 60637, USA}
\affiliation{$^{2}$Kavli Institute for Cosmological Physics, University of Chicago, Chicago IL 60637, USA}
\affiliation{$^{3}$Sub-department of Astrophysics, University of Oxford, Keble Road, Oxford OX1 3RH, United Kingdom}
\affiliation{$^{4}$Institut d'Astrophysique de Paris, Sorbonne Universites, CNRS, UMR 7095, 98 bis bd Arago, 75014 Paris, France}
\affiliation{$^{5}$University of Bath, Department of Physics, Claverton Down, Bath, BA2 7AY, UK}
\affiliation{$^{6}$Universite Claude Bernard Lyon 1, CRAL UMR5574, ENS de Lyon, CNRS, Villeurbanne, F-69622, France}
\affiliation{$^{7}$Cosmic Dawn Center (DAWN), Copenhagen, Denmark} 
\affiliation{$^{8}$Niels Bohr Institute, University of Copenhagen, Jagtvej 128, DK-2200, Copenhagen, Denmark}
\affiliation{$^{9}$Institute for Astronomy, University of Edinburgh, Royal Observatory, Edinburgh EH9 3HJ, UK}
\affiliation{$^{10}$Department of Astronomy, Yonsei University, 50 Yonsei-ro, Seodaemun-gu, Seoul 03722, Republic of Korea}
\affiliation{$^{11}$Sterrenkundig Observatorium Department of Physics and Astronomy Universiteit Gent, Krijgslaan 281 S9, B-9000 Gent, Belgium}

\begin{abstract}
The ultraviolet (UV) spectral slope, $\beta$, serves as a fundamental probe of stellar population properties and interstellar medium conditions in galaxies. At high redshift, a small sample of galaxies have been confirmed to have extremely blue spectral slopes of $\beta<-2.8$, as measured directly from spectroscopy. One explanation for such galaxies is that they are leaking copious amounts of ionizing photons such that these photons are never processed into the nebular continuum and we are observing the bluer intrinsic slopes of the stellar populations. Here, we use a detailed set of radiation hydrodynamics simulations with parsec-scale resolution and non-equilibrium chemistry to elucidate the physical origin of extremely blue UV slopes at high redshift. We show that, while some extremely blue galaxies (EBGs) do indeed have high escape fractions, a second population of EBGs has $f_{\rm esc}<1\%$. In this second population, the ISM gas densities tend to be much lower, and the timescale for ionization fronts to reach their Strömgren radii can be longer than the main-sequence lifetime of massive stars, leading to a lag in nebular emission. Our results demonstrate that extremely blue UV slopes do not uniquely imply high escape fractions, which highlights the importance of time-dependent (non-equilibrium) nebular emission for interpreting the spectra of galaxies in the early Universe.
\end{abstract}

\section{Introduction}
The ultraviolet spectral slope ($\beta$) represents a key diagnostic of the stellar and gaseous properties of galaxies \citep[e.g.][]{Bouwens2010,Mclure2011,Finkelstein2012,Dunlop2013,Wilkins2016,Narayanan2025}. $\beta$ is particularly useful at extremely high-redshifts where typically only the rest-frame UV of a galaxy spectral energy distribution (SED) is accessible to space-based telescopes such as HST and JWST\footnote{This primarily applies to NIRSpec and excludes the few sources that have been detected with MIRI \citep[e.g.][]{Zavala2025,Marquez2025,Helton2025}. Likewise, there have also been a few single emission line detections with ALMA \citep[e.g.][]{Carniani2025}. Nevertheless, these detections represent only a small fraction of the spectroscopically confirmed $z>9$ galaxy population.}. Moreover, because $\beta$ can be computed from both spectra and photometry, large samples ($>$1,000) of robust UV slope measurements have been made at $z>6$ \citep[e.g.][]{Cullen2024,Topping2024,Saxena2024,RB2024,Tang2025}.

For a simple stellar population (SSP), in the absence of dust, UV slope traces two key properties: the age of the stellar population, and the strength of the nebular continuum. The most massive stars in a population are both the brightest and the bluest. As a result, the intrinsic UV slope of a stellar population tends to remain constant until the first massive stars evolve off the main-sequence, before becoming subsequently redder as the lower-mass stars further evolve. Hence the intrinsic UV slopes of an SSP are expected to redden with time. (see e.g.\ Figure~6 of \citealt{Katz2024_BJ}). 

However, the most massive stars are also the most efficient producers of ionizing photons. A fraction of these photons are absorbed by the surrounding gas, ionizing hydrogen atoms, with the now-free electrons driving nebular continuum emission through free-free, free-bound and two-photon processes. Nebular continuum emission tends to redden the observed spectrum, primarily due to the fact that the free-bound emission increases strongly towards longer wavelengths in the interval $912-3645$~\AA. Hence, for commonly assumed SSPs and ISM properties\footnote{Assuming a typical stellar population with a Kroupa \citep{Kroupa2001} IMF and a maximum mass of 100~M$_{\odot}$.}, $\beta$ tends to remain constant at a value of $\sim-2.6$ for the first 10~Myr and then reddens as the nebular continuum weakens and the SSP ages (see e.g.\ Figure~9 of \citealt{Cullen2024}). 

There are numerous physical effects that complicate the simple $\beta\sim-2.6$ scenario described above. First, real galaxies are not SSPs but rather exhibit complex star formation histories. Any additional contribution from older stellar populations will redden the UV slope \citep[e.g.][]{Narayanan2024}, rather than making it bluer. Similarly, the magnitude and wavelength-dependence of dust reddening depend sensitively on the microphysical properties of the dust grains and covering fractions, which may be different at high redshift \citep[e.g.][]{Reddy2025,McKinney2025} compared to commonly assumed dust attenuation curves \citep[e.g.][]{Calzetti2000}. However, the effect of dust is always to redden the UV slope.

To make galaxies UV-bluer, one could imagine modifying the properties of the stellar populations. A more top-heavy IMF where one simply flattens the high mass slope without changing the upper mass limit, for example, makes the intrinsic UV slopes bluer. However, the contribution from the nebular continuum also increases, and so the net effect is that the observed spectra are mildly redder \citep[e.g.][]{Katz2024_BJ}. Similarly, allowing for the formation of very massive stars with $M>100$~M$_{\odot}$ (which have intrinsic UV slopes bluer than stars of 100~M$_{\odot}$) also leads to enhanced ionizing photon production, and in turn increases the nebular contribution, which further reddens the spectrum \citep[e.g.][]{Schaerer2025}. The same effect causes Pop.~III stars embedded in dense gas to appear redder than star-forming Pop.~II systems \citep[e.g.][]{Schaerer2002,Inoue2011,Zackrisson2011,Trussler2023,Katz2025-meg}. 

However, if the ionizing photons are allowed to escape the ISM without interactions, no nebular continuum emission is produced and the observed spectrum may be closer to the intrinsic values of the stellar population. Such an effect has been postulated from photoionization models \citep[e.g.][]{Zackrisson2013} and numerical simulations \citep[e.g.][]{Choustikov2024}, and observed in a subset of local reionization analog galaxies \citep{Chisholm2022}. 

However, a simple link between $\beta$ and $f_{\rm esc}$ may not be fully representative of the high-redshift galaxy population. Adopting a scaling between UV slope and $f_{\rm esc}$ causes 1D models of reionization to systematically reionize too early \citep{Munoz2024}. Furthermore, galaxies shown in \cite{Saxena2024} not only have extremely blue UV slopes, but also high equivalent widths emission lines, notably H$\alpha$ and H$\beta$. The latter characteristic, at face value, seems incompatible with the idea that ionizing photons are escaping since hydrogen recombination lines are powered by ionizing photons being absorbed by ISM gas. While it is very likely that some of the galaxies with extremely blue UV slopes are leaking ionizing photons (see e.g.\ discussion in \citealt{Yanagisawa2025,Cullen2025}), it is possible that there is some other physical mechanism driving the behaviour seen for some galaxies in \cite{Saxena2024}.

In this work, we search the {\small MEGATRON} suite of high-redshift galaxy formation simulations for extremely-blue galaxies. This suite varies key assumptions in star formation, stellar evolution and feedback modelling, allowing us to compare the spectral properties of galaxy populations across models (see \citealt{Katz2025-meg} for further details). We identify extremely-blue galaxies and highlight the key physical characteristics that leads to steep $\beta$. 

\vspace{2cm}
\section{The physical origin of extremely blue $\beta$}
Our goal is to understand how and why certain galaxies exhibit extremely blue UV slopes. Here we adopt a conservative approach and define an extremely blue galaxy as one that has $\beta < -2.8$. 

\begin{figure*}
  \centering
    \includegraphics[width=\textwidth]{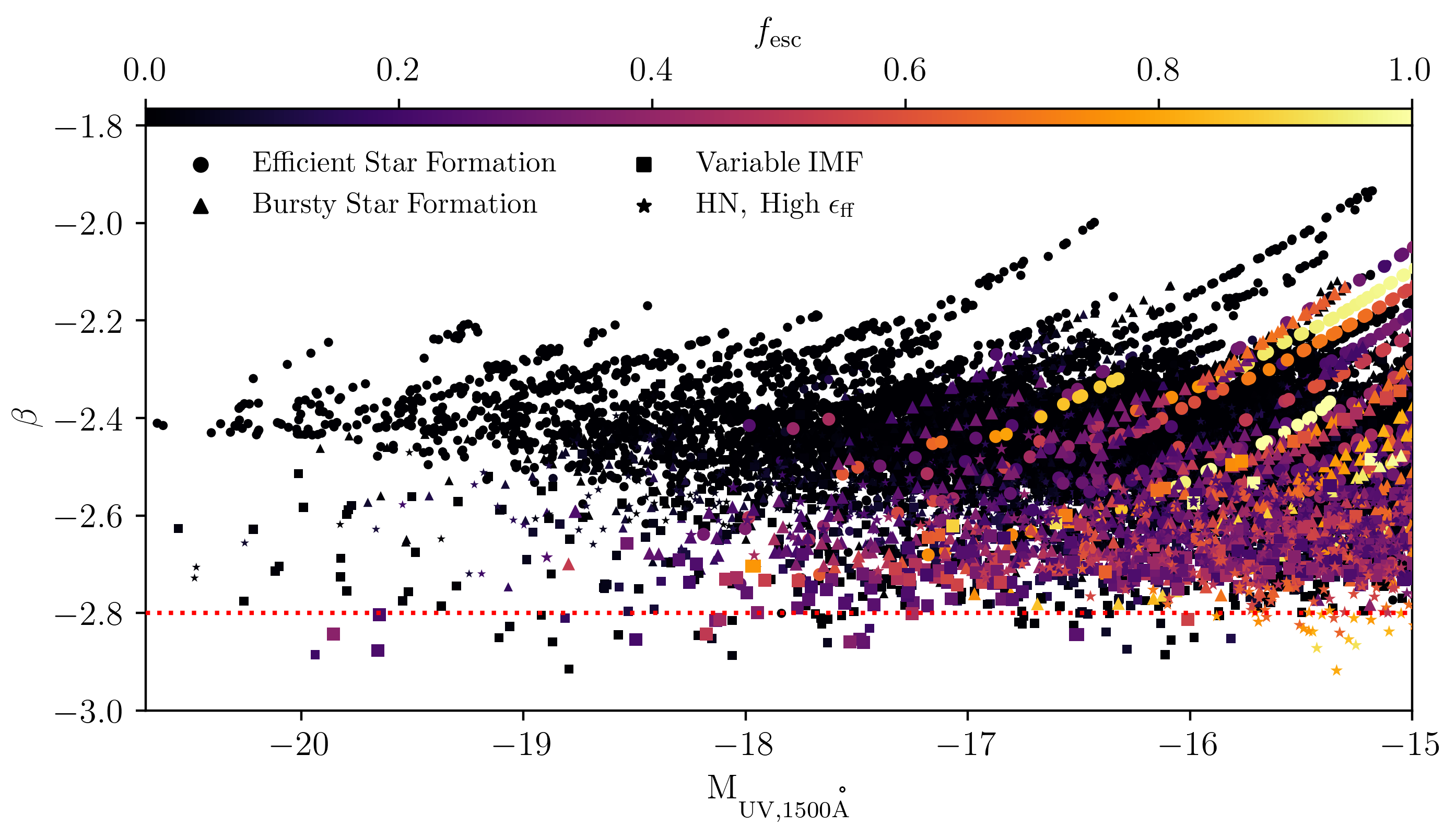}
    \includegraphics[width=\textwidth]{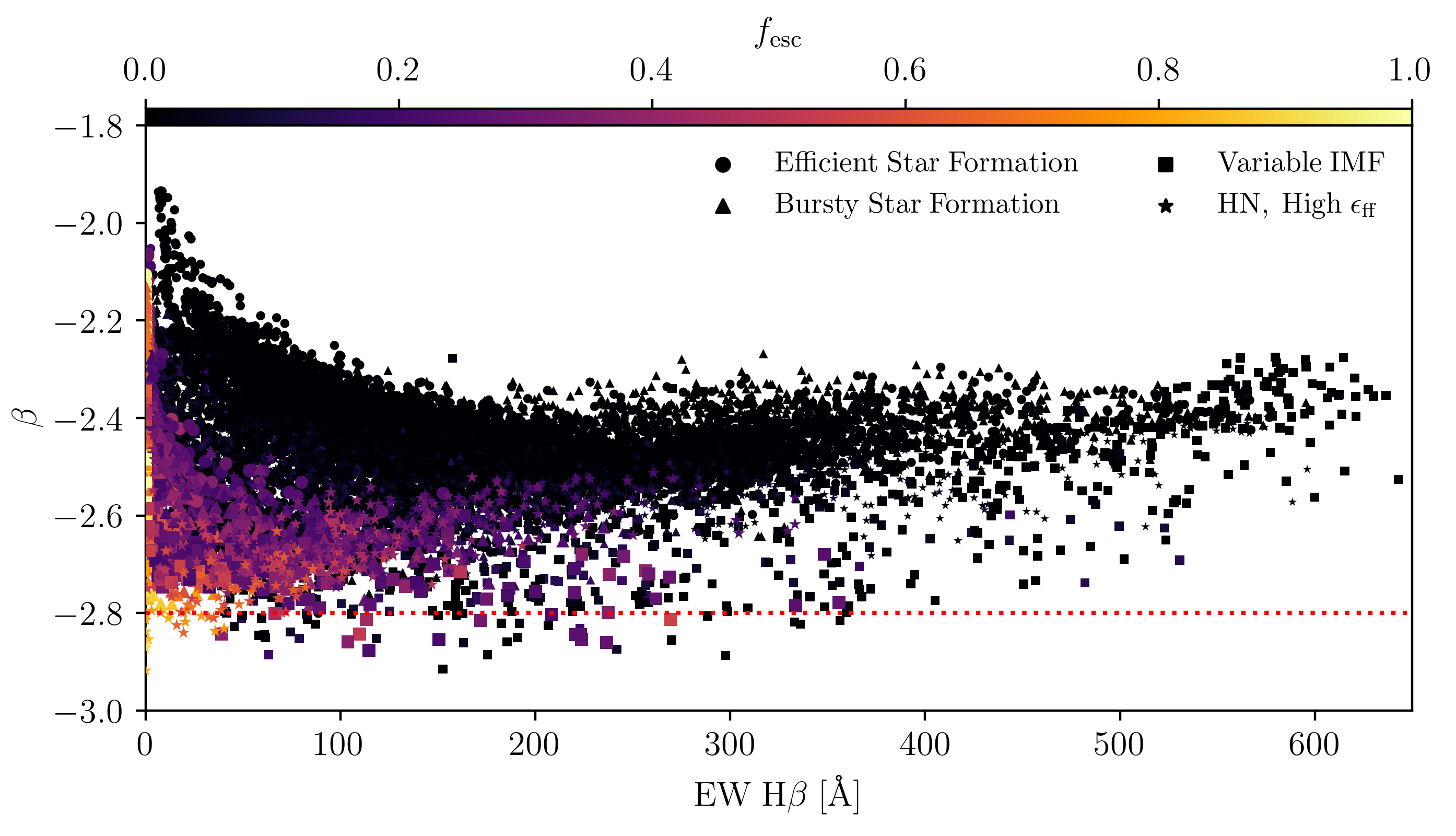}
    \caption{UV slope as a function of 1500~\AA\ UV magnitude (top) or H$\beta$ equivalent width (bottom) for all high-redshift MEGATRON simulations (as indicated in the legend). The spectra include both the stellar and nebular continua.  We colour each point by the LyC escape fraction measured at 900~\AA\ and consider only central haloes (i.e.\ ignoring sub-haloes). For guidance, we plot a horizontal dashed red line at $\beta=-2.8$ which represents our conservative definition of an extremely blue galaxy. Tracks represent individual galaxies evolving over time. Many of the galaxies with steep UV slopes exhibit low $f_{\rm esc}$ and high H$\beta$ equivalent widths. }
    \label{fig:muv_beta} 
\end{figure*}

\subsection{Numerical Simulations}
In this work, we utilize the high-redshift suite of {\small MEGATRON} simulations \cite{Katz2024_meg,Katz2025-meg,Rey2025,Cadiou2025}. These simulations are ideal for our purposes as they reach very high spatial resolution ($\lesssim 5$~pc) and a detailed non-equilibrium chemistry network consisting of primordial species, metals, and molecules is coupled to multi-frequency radiation transport, allowing the spectra of the simulated galaxies to be predicted by the simulations. Full details of these simulations can be found in \cite{Katz2024_meg} and \cite{Katz2025-meg}. The important aspect for this work is that we study four realizations of the same initial conditions that sample different subgrid models for star formation, feedback and stellar populations. More specifically the models are:
\begin{enumerate}
    \item Efficient star formation: this ``vanilla'' model exhibits the highest conversion efficiency of gas into stars. The feedback is the weakest among the four models and so star formation remains relatively unregulated, leading to very bright galaxies at early times. Stellar SEDs follow BPASS \citep{Eldridge2017,Stanway2018} assuming a Kroupa IMF with a maximum mass of 300~M$_{\odot}$.
    \item Bursty star formation: this model increases the energy per supernova by a factor of five compared to the efficient star formation model. Galaxy stellar masses are better regulated but the star formation histories become very ``bursty''. Stellar SEDs follow BPASS \citep{Eldridge2017,Stanway2018} assuming a Kroupa IMF with a maximum mass of 300~M$_{\odot}$.
    \item Variable IMF: this model adopts a density and metallicity dependence for the upper mass slope of the stellar IMF following \cite{Marks2012}. Furthermore, a subset of high mass stars explode as hypernovae, releasing $10-40\times10^{51}$~ergs per event. The upper mass slope of the IMF can vary, but the upper mass limit remains fixed at 120~M$_{\odot}$. The stellar SEDs follow Starburst99 \citep{Leitherer1999}.
    \item HN, High $\epsilon_{\rm ff}$: the final model forms stars in lower density gas at much higher efficiency, adopting a rate of 100\% per free-fall time. This leads to bursty star formation histories and a slightly lower density ISM. Furthermore, like the IMF model, a subset of high mass stars explode as hypernovae. Stellar SEDs follow BPASS \citep{Eldridge2017,Stanway2018} assuming a Kroupa IMF with a maximum mass of 300~M$_{\odot}$. This model best regulates the stellar mass.  
\end{enumerate}

For each simulation, we produce a library of $40,000-50,000$ spectra considering the stellar continuum, nebular continuum, and nebular emission lines for a total of $>175,000$ spectra. We include all galaxies at $z\geq8.5$ --- the final redshift of the simulations. Note that the spectral library includes the same galaxies at multiple different times. Depending on the simulation, in the final snapshot, the number of unique galaxies that are resolved enough to be included in our spectral database ranges from $\approx$ 400 to 600. Outputs are generated for each simulation at least every 5~Myr. This cadence can be finer in practice if the simulation checkpointed in between two planned outputs when finishing a run cycle. Even though our output cadence is high, SFRs can fluctuate on Myr timescales while ionization and recombination timescales in the high-redshift ISM are $\approx 1-3$~Myr. Thus there is a chance that we miss some galaxies that exhibit the non-equilibrium physics we are interested in between snapshots.

We ignore the role of dust here for two reasons. First, we are primarily interested in low-mass galaxies, which are less chemically evolved and thus have lower dust content. This is more consistent with observations that show minimal dust attenuation at these redshifts \citep[e.g.][]{Topping2024,Cullen2024}. Second, dust reddens the spectrum. If our simulations cannot produce galaxies with extremely blue UV slopes without dust, including it would only further aggravate the disagreement.

UV slopes are calculated following \cite{Saxena2024} from the simulated spectra in the wavelength interval 1,350$-$2,700~\AA. We include the nebular continuum when fitting for $\beta$, but exclude nebular emission lines. Escape fractions in this work are monochromatic, corresponding to the value at 900~\AA. We measure the escape fraction at 75\% of $R_{\rm vir}$, and the spectrum of each galaxy only considers gas and stars within 25\% of $R_{\rm vir}$. The impact of these choices is discussed in Appendix~\ref{app:aperture}.

\subsection{Results}
We begin by showing how the UV slope varies with 1,500~\AA\ UV magnitude and H$\beta$ equivalent width (Figure~\ref{fig:muv_beta}). We have coloured each point by the LyC escape fraction of the galaxy and only consider central galaxies with M$_{\rm UV} < -15$. The horizontal red dotted line represents our demarcation for an extremely blue galaxy (EBG) at $\beta=-2.8$. 

Unsurprisingly, the vast majority of galaxies in the {\small MEGATRON} simulations are not EBGs and have low $f_{\rm esc}$.  This is consistent with previous simulations \citep[e.g.][]{Rosdahl2018,spdrv1} and observations of galaxies in the high-redshift \citep[e.g.][]{Saxena2024,Cullen2024,Topping2024,Austin2024}. Among the EBG population, there is a continuum of sources including galaxies with very high $f_{\rm esc}$ and those with $f_{\rm esc}\sim0$.

\begin{figure}
  \centering
    \includegraphics[width=\columnwidth]{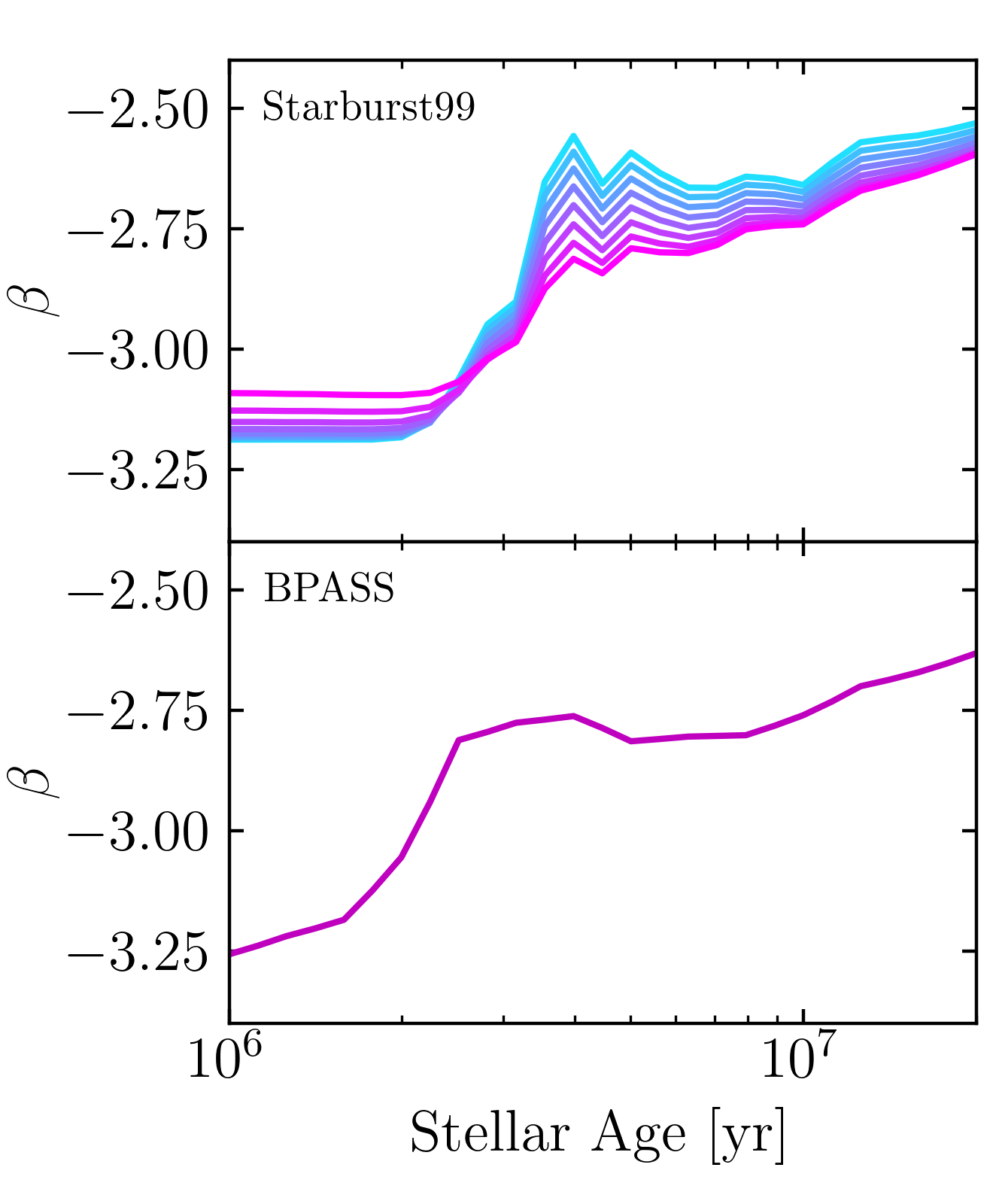}
    \caption{Intrinsic UV slopes of stellar populations (i.e.\ those measured without the contribution of the nebular continuum) from Starburst99 (top) and BPASS (bottom) as a function of stellar age. We adopt the same Starburst99 and BPASS parameters as used in the MEGATRON simulations. For demonstrative purposes, we only show SSP models with $Z=0.01Z_{\odot}$. The colour of the line represents the slope of the IMF ranging from $-0.5$ to $-2.6$ from cyan to magenta. We consider only one IMF slope for the BPASS models.}
    \label{fig:intrinsic_beta}
\end{figure}

The galaxies that exhibit high $f_{\rm esc}$ and blue $\beta$ are expected from both theory and observation \citep[e.g.][]{Chisholm2022,Papovich2025,Giovinazzo2025}. The intrinsic UV slopes of {\small BPASS} and {\small Starburst99} SSP models are $<-3$ at ages $\lesssim2$~Myr (see Figure~\ref{fig:intrinsic_beta}). When there is little gas in the system, the observed spectrum tends towards the intrinsic stellar continuum. The galaxies with $f_{\rm esc}\sim100\%$ and blue $\beta$ cluster at faint UV magnitudes with M$_{\rm UV}\lesssim-16$. There are two reasons why this occurs. First, to have such high $f_{\rm esc}$ and very steep $\beta$ the galaxy must be nearly gas-free. It is easier for feedback to expel the gas from low-mass haloes  \citep[e.g.][]{Dekel1986} and thus faint galaxies may be more susceptible to such strong catastrophic feedback events. Such behaviour is seen in other simulations of high-redshift galaxies \citep[e.g.][]{Xu2016b,Rosdahl2018,Ma2018}. Note that we also have a population of galaxies with high $f_{\rm esc}\sim100\%$ but not extreme UV slopes. These represent ``remnant leakers'' (as discussed in \citealt{Katz2023_bursty}) and are likely the progenitors of mini-quenched galaxies \citep[e.g.][]{Looser2024,Strait2023,Trussler2025,Baker2025}. While they have high $f_{\rm esc}$, their stellar populations are older and they do not emit many ionizing photons. The second reason that EBGs with high $f_{\rm esc}$ cluster at fainter UV magnitudes is that emission from older stellar populations due to an extended star formation history can also redden $\beta$. High specific SFRs (sSFR) are needed for blue UV slopes, which occur more frequently in lower mass haloes.

To reach such high values of $f_{\rm esc}$ so that the nebular continuum is weak enough to form an EBG, the feedback must occur early and be very strong. Hence, most of the EBGs in our simulations with high $f_{\rm esc}$ come from the HN, High $\epsilon_{\rm ff}$ and Variable IMF simulations. Both of these simulations include hypernova feedback, which occurs at earlier times in the life-cycle of a stellar population (i.e.\ when $\beta$ is bluest) compared to a normal core-collapse supernova. Furthermore, the additional energy injection from a hypernova increases the probability of unbinding the gas. 

We also find EBGs with $f_{\rm esc}\sim0$. The existence of such a population of galaxies is perhaps surprising. It is clear from Figure~\ref{fig:intrinsic_beta} that to have an intrinsic $\beta\leq-2.8$, the age of the stellar population must be less than a few Myr. This is because the most massive stars have the steepest UV slopes. In the extreme limit of only very massive stars with $T\sim10^5$~K, the UV slope is limited to a value of $\sim-3.6$ (as measured from a blackbody). However, these massive stars also emit the most ionizing photons, which are reprocessed by the nebula so that the observed UV slopes are typically redder than stellar populations without hot massive stars \citep{Katz2024_BJ}. This explains why SSP models with a top-heavy IMF have bluer intrinsic UV slopes but redder observed slopes when $f_{\rm esc}\sim0$. The fact that any galaxies in our simulations have $f_{\rm esc}\sim0$ and $\beta<-2.8$ is extremely puzzling.  

\begin{figure}
  \centering
    \includegraphics[width=\columnwidth]{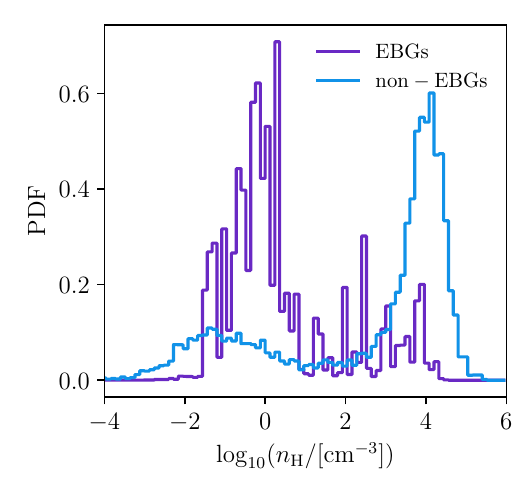}
    \caption{Stacked PDFs of gas density around star particles weighted by their LyC luminosity from the variable IMF simulation. Purple represents our EBG sample with $M_{\rm UV}<-15$, $f_{\rm esc}<1\%$, and $\beta<-2.8$, while blue represents the non-EBG comparison sample with $M_{\rm UV}<-15$, $f_{\rm esc}<1\%$, and $-2.8<\beta<-2.5$. The mode of the density PDF of non-EBGs galls at very high density, while for the EBGs, the mode is at very low density. Note that the EBG histogram is noisier than that for the non-EBGs due to the smaller sample size (24 vs. 408 galaxies).}
    \label{fig:ebg_dens}
\end{figure}

To understand the physical origin of the extremely blue slopes with simultaneous low $f_{\rm esc}$, for each galaxy, we calculate the probability distribution (PDF) of gas density for cells in each galaxy weighted by the amount of ionizing photons currently being injected into each gas cell by the star particles. We compare stacked PDFs of bright EBGs with low $f_{\rm esc}$ with a similarly bright sample of non-EBGs with low $f_{\rm esc}$ in Figure~\ref{fig:ebg_dens}. More specifically, we select all main haloes with M$_{\rm UV}<-15$, $f_{\rm esc}<1\%$, and $\beta<-2.8$ as the EBG sample and identify a comparison sample of non-EBGs with M$_{\rm UV}<-15$, $f_{\rm esc}<1\%$, and $-2.8<\beta<-2.5$. The mode of the PDF for the non-EBGs is at a gas density that is $>10^4\times$ higher than the EBGs.  

In principle, $\beta$ is insensitive to gas density\footnote{This is true as long as the density remains that where the $2s$ state of neutral hydrogen is not collisionally de-excited. Most of the non-EBGs have H~{\tiny II} region gas densities above this critical value; nevertheless, the UV slope is primarily driven by the free-bound emission rather than the two-photon emission. Collisionally de-exciting the two-photon emission can often make the SED redder.}. To understand this, consider a fully ionized gas of pure hydrogen. The luminosity of the continuum emission (from recombination) goes as
\begin{equation}
    L_{\lambda} = \epsilon(T,\lambda)n_{\rm H}^2 V,
\end{equation}
where $\epsilon(T,\lambda)$ is the emissivity of the continuum as a function of wavelength and temperature, $n_{\rm H}$ is the hydrogen gas density, and $V$ is the volume of the ionized gas. The volume can be computed as
\begin{equation}
    V = \frac{4}{3}\pi R_s^3,
\end{equation}
where the Strömgren radius is
\begin{equation}
    R_s = \left(\frac{3}{4\pi}\frac{Q}{\alpha_B(T)n_{\rm H}^2}\right)^{1/3},
\end{equation}
and $Q$ is the emission rate of ionizing photons per second and $\alpha_B(T)$ is the Case~B recombination rate. Substituting for volume, we find that
\begin{equation}
    L_{\lambda} = \frac{\epsilon(T,\lambda)Q}{\alpha_B(T)}.
\end{equation}
Thus even if the ionizing photons are leaking from the ISM, as long as they are absorbed at some point in the halo (which is the case when $f_{\rm esc}\sim0$), the gas density has no immediate impact on the nebular contribution to the SED. There remains a weak dependence on temperature, but this is a small effect. Furthermore, cooling radiation can contribute non-negligibly to the two-photon emission, which is not considered here, but will matter when the nebula is sufficiently hot \citep[e.g.][]{Raiter2010,Katz2024_BJ}.

However, the above derivation only applies in the equilibrium limit, i.e.\ when the Strömgren sphere has been fully developed and the system is in steady-state. Such an assumption is the default in photoionization models. However, it remains unclear whether these systems have actually reached steady-state. To address this, we can calculate the timescale required for the ionization front to reach $R_s$. 

\begin{figure}
  \centering
    \includegraphics[width=\columnwidth]{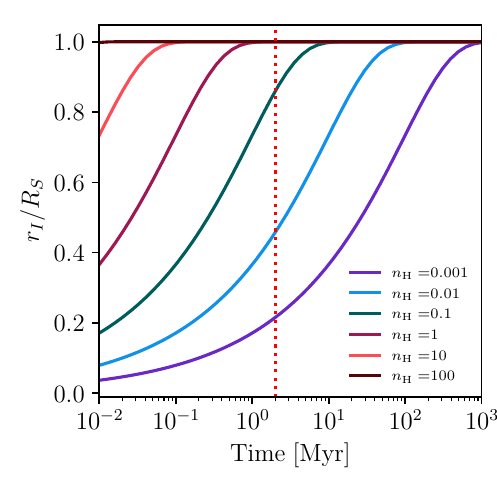}
    \caption{Ratio of the ionization front radius to the Strömgren radius as a function of time for various gas densities (colours). We assume a fiducial temperature of $T=2\times10^4$~K. The vertical red lines highlight the timescale of $2$~Myr which are the typical LyC weighted ages of bright EBGs in the variable IMF simulation.}
    \label{fig:r_time}
\end{figure}

Following \cite{Spitzer1978,Iliev2006}, the analytical solution for the ionization front radius as a function of time is given as
\begin{equation}
    r_I = R_s (1-e^{t/t_{\rm rec}})^{1/3},
\end{equation}
where $t_{\rm rec} = 1/\alpha_B(T)n_{\rm H}$ is the recombination timescale. As one can see, the ratio of $r_I/R_s$ depends only on gas density and mildly on gas temperature\footnote{Note that in the simulations, we use a reduced speed of light approximation. The impact of this approximation is discussed in Appendix~\ref{app:rsla}.}. Taking a fiducial value of $T=2\times10^4$~K, in Figure~\ref{fig:r_time} we show $r_I/R_s$ as a function of time for various gas densities. We highlight the timescale of $2$~Myr, which is when the intrinsic UV slopes of our adopted SSP models begin to redden (see Figure~\ref{fig:intrinsic_beta}). We find that for a gas density 0.1~cm$^{-3}$, the ionization front only reaches $86\%$ of the Strömgren radius at $2$~Myr. Since $L\propto r_I^3$, we can expect the nebular continuum to be weaker by a factor of $0.63$. Of course, the radiation can travel further away than the immediate cell of the star particle and the gas becomes less dense the further it travels into the halo. Hence, if the radiation is able to escape the immediate ISM surrounding the star particle and into the much lower density circumgalactic medium (CGM), timescales begin to matter for the nebular emission. In other words, the system is out of equilibrium, which has the effect of reducing the nebular emission. Thus, we have identified a mechanism which leads to steep UV slopes at low $f_{\rm esc}$. The condition for this to occur is $r_I/R_S < 1$ at $t\lesssim 2$~Myr and $R_s<R_{\rm vir}$, where $R_{\rm vir}$ is the virial radius of the halo.

Stated another way, the Strömgren radius is reached and the system is in equilibrium when the number of absorptions is balanced by the number of recombinations. In this limit, every ionizing photon can be (instantaneously) paired with a recombining atom so that the luminosity is only proportional to $Q$. However, simply analyzing the recombination timescale demonstrates that this assumption may not hold if photons are emitted into low density gas. The recombination rate for H~{\small II} assuming Case~B can be approximated as $\alpha(T)=2.753\times10^{-14}\lambda_{\rm HI}^{1.5}(1+(\lambda_{\rm HI}/2.74)^{0.407})^{-2.242}\ {\rm cm^3\ s^{-1}}$, where $\lambda_{\rm HI}=315614\ {\rm K} / T$ \citep{Hui1997}. Adopting a fiducial value of $2\times10^4$~K, $\alpha=1.43\times10^{-13}\ {\rm cm^3\ s^{-1}}$. At an electron density of $n_e=0.1\ {\rm cm^{-3}}$, the recombination timescale is 3~Myr, which is similar to the main-sequence lifetimes of massive stars. As the ionizing photons are emitted from the star, they ionize the gas, but the recombination event, which gives rise to the nebular continuum, only occurs approximately 3~Myr later. Hence there is a time delay between emission and absorption that impacts the observed spectrum.

\begin{figure}
  \centering
    \includegraphics[width=\columnwidth]{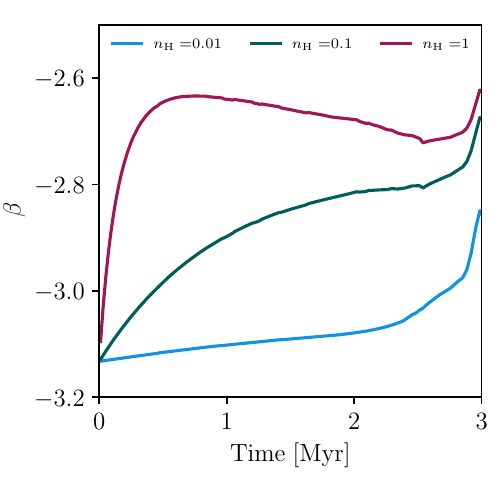}
    \includegraphics[width=\columnwidth]{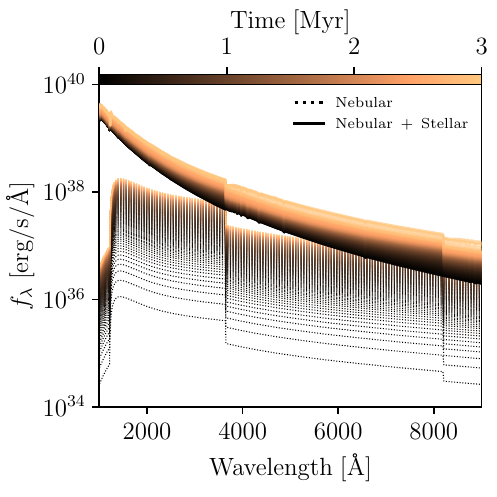}
    \caption{(Top) UV slope as a function of time for a {\small STARBURST99} model SSP with a mass of $10^6\ {\rm M_{\odot}}$, a Kroupa IMF, a maximum mass of 120~M$_{\odot}$, and with a metallicity of 2\%~$Z_{\odot}$, irradiating gas with density of 0.01, 0.1, or 1~${\rm cm^{-3}}$. For the lower density models, $\beta$ slowly reddens with time as the ionization front does not reach the Strömgren radius within the lifetimes of the most massive stars. (Bottom) Example spectra at different times for the model with an ISM density of 0.1~${\rm cm^{-3}}$. As time progresses, the nebular continuum contributes more to the total spectrum and the UV slope reddens.}
    \label{fig:beta_time}
\end{figure}

To demonstrate the impact of this non-equilibrium effect on $\beta$, we calculate how the UV slope might evolve as a function of time for an SSP irradiating gas of different densities. For our experiment, we consider a {\small STARBURST99} model SSP with a mass of $10^6\ {\rm M_{\odot}}$, a Kroupa IMF, a maximum mass of 120~M$_{\odot}$, and with a metallicity of 2\%~$Z_{\odot}$, irradiating gas with hydrogen number density ($n_{\rm H}$) of 0.01, 0.1, or 1~${\rm cm^{-3}}$. For simplicity, we assume the temperature of the gas is always $2\times10^4$~K and that $Q$ is the mean ionizing luminosity of the SSP between $1-3$~Myr\footnote{In practice, the ionizing luminosity only deviates by a maximum of 8\% during this time period and so this is a very good approximation.}. The nebular continuum contribution to the total SED is computed with {\small PyNeb} \citep{Luridiana2015}. The top panel of Figure~\ref{fig:beta_time} shows how $\beta$ evolves as a function of time for each of the ambient gas densities while the bottom panel shows the full spectrum in the UV and optical as a function of time for the model with $n_{\rm H}=0.1\ {\rm cm^{-3}}$. The model with $n_{\rm H}=1\ {\rm cm^{-3}}$ plateaus at a value $\beta\sim-2.7$ in $<1$~Myr. This is consistent with the expectations from Figure~\ref{fig:r_time} where the ionization front reaches the Strömgren radius within this timescale. In contrast, the other two models never plateau and $\beta$ simply reddens with time as the ionization front expands. $\beta$ is always $<-2.8$ in these lower density models over the lifetime of the massive stars, which naturally explains the EBGs in our simulation that have $f_{\rm esc}\sim0$.

\section{Discussion \& Conclusions}
To address the question of what is the physical origin of steep UV slopes at high redshift, we find two physical mechanisms: 1) high LyC escape fraction and 2) non-equilibrium effects. 

The former has been discussed extensively in the literature, and there exist numerous calibrations, both theoretical and empirical \citep[e.g.][]{Zackrisson2013,Chisholm2022,Choustikov2024} as well as applications of such models to observed high-redshift galaxies \citep[e.g.][]{Papovich2025,Giovinazzo2025}. Our numerical simulations form a subset of EBGs with very high $f_{\rm esc}$, approaching 100\%, which clearly highlights this mechanism at work. It is important to consider the fact that the reddening is sensitive to $1-f_{\rm esc}$. Astrophysically relevant values of $f_{\rm esc}$ that can lead to a complete reionization by $z\sim5.3$ are on the order of 10\% \citep[e.g.][]{Robertson2015,Rosdahl2018,Finkelstein2019} or possibly lower and thus the Universe can almost certainly reionize with a negligible impact of $f_{\rm esc}$ on the observed UV slopes. Reaching escape fractions of 100\% is rare in our simulations and typically happens after supernova feedback, at which point the massive stars have evolved off the main-sequence and thus the intrinsic UV slope of the stellar population is redder. Nevertheless, this does occur, particularly in lower mass galaxies in the simulations that are more susceptible to feedback \citep[e.g.][]{Dekel1986}. 

Even in the case of high $f_{\rm esc}$, to have a UV slope of $\beta\lesssim-2.8$, a requirement is that the galaxies have very bursty star formation histories. This is so that the young stars can completely outshine the older stellar populations so that the older stars do not redden the spectrum. The amount of variation in UV magnitude of galaxies is currently hotly debated \citep[e.g.][]{Pallottini2023,Sun2024}, although almost all simulations predict a strong mass dependence such that more massive galaxies are less ``bursty'' \citep[e.g.][]{Gelli2024a,Kravtsov2024,Shen2025}. This mass dependence further biases steep UV slopes to lower mass galaxies. In this regard, a top-heavy IMF (defined as one with a fixed upper mass limit but a shallower slope) can help produce EBGs at brighter magnitudes. This is because stellar populations with a top-heavy IMF have a lower mass-to-light ratio at early times, and so ``outshining'' becomes easier. Likewise, at later times, the SEDs are less bright because more mass has evolved off the main sequence. Hence the contribution from older stellar populations reddens the spectra less than it would for a more canonical IMF. When $f_{\rm esc}=0$, the SEDs of SSPs with a top-heavy IMF are expected to be redder than for a normal IMF \citep[e.g.][or Figure~\ref{fig:beta_time}]{Katz2024_BJ} and so when accounting for the possibility of a non-zero $f_{\rm esc}$, we expect an increase in the scatter in $\beta$ at all magnitudes.

In contrast to the scenario of high $f_{\rm esc}$, the non-equilibrium mechanism has rarely been discussed in the literature (if at all). The key effect is that if the timescale for the ionization front to reach the Strömgren radius (or put another way, the recombination timescale) is on the order of or longer than the main-sequence lifetime of massive stars, the nebular continuum will appear weaker than what one obtains from an equilibrium photoionization model. Furthermore, if the Strömgren radius is smaller than the virial radius of the halo, no ionizing photons escape or contribute to the reionization, thus leading to a scenario where $\beta$ is steep but $f_{\rm esc}\sim0$. 

To satisfy the timescale constraint, the ionizing photons must be present in gas with $n_{\rm H}\lesssim0.1\ {\rm cm^{-3}}$. This density is significantly lower than that observed at high-redshift \citep[e.g.][]{Isobe2023,Topping2025}; however, some fraction of the ionizing photons may be escaping through channels into lower density gas. In our simulations, particularly the ones with the variable IMF, we find that stellar birth clouds can be efficiently disrupted and a large fraction of the ionizing photons are emitted into gas with densities low enough such that non-equilibrium effects become important (see Figure~\ref{fig:ebg_dens}). Early stellar feedback is necessary to form EBGs, unless the turbulence in the cloud is so high that there are naturally optically thin channels for radiation to escape. 

From an observational perspective, there is good reason to believe that star-forming clouds are quickly disrupted. The spatial de-correlation between ionized gas and molecular gas \citep[e.g.][]{Kruijssen2019,Ramambason2025} observed in local galaxies seems to require efficient early feedback \citep[e.g.][]{Semenov2021}. High-resolution simulations of turbulent giant molecular clouds also predict that radiation feedback disrupts them well before massive stars evolve off the main-sequence \citep[e.g.][]{Kimm2022,Menon2024}. Observations of local LyC reionization analogs also support the picture of pre-SN feedback creating channels or completely disrupting birth clouds \citep{Flury2025,Komarova2025,Carr2025}. At high redshifts, lower metallicities (and presumably dust content) render IR radiation pressure less effective and likewise, stellar winds are also less important \citep[e.g.][]{Jecmen2023}. In contrast, the increased ionizing emissivity of low-metallicity massive stars compared to solar metallicity stars and Ly$\alpha$ radiation pressure may become much more important \citep{Kimm2018,Nebrin2025}. 

It is unclear if the subgrid physics in the {\small MEGATRON} simulations is disrupting the GMCs for the correct physical reasons; however, assuming some portion of galaxies are able to efficiently disrupt a significant fraction of the stellar birth clouds or large ionized channels are produced, then non-equilibrium effects may become important as shown here. We emphasize that even in the variable IMF simulation, which produces the most EBGs, EBGs are rare, representing $<2\%$ of all galaxies with ${\rm M}_{\rm UV}<-15$. Even if our simulations are not a completely accurate picture of reality, they highlight how if such ISM conditions are realized in the real Universe, then one might have to consider non-equilibrium effects when interpreting certain high-redshift SEDs.

In our current analysis, we have neglected the impact of dust on the mock spectra of the simulated galaxies. This omission prohibits a detailed comparison between the observed distribution of UV slopes as a function of UV magnitude with the simulation data; however, as discussed above, dust only acts to redden the spectra. Thus, if our models cannot produce EBGs without dust, including its effects would further worsen the comparison between simulations and observations.

It is important to consider that our simulations may be missing some physical effects that could prevent non-equilibrium effects from becoming important more often. First, H~{\small II} regions are not always well resolved. In particular, their detailed turbulent structure and the ionized channels that may form from early stellar feedback are not always well captured. Turbulence can broaden the gas density PDF and create additional optically thin channels for radiation to leak into the lower-density ISM and CGM  \citep[e.g.][]{Kakiichi2021}. In this analysis, when the Strömgren sphere is not resolved, we replace the emission from the gas cell (nebular continuum and emission lines) with an equilibrium solution. This maximizes the nebular contribution to the total SED which may cause us to underestimate the non-equilibrium effects that may be present if the turbulent H~{\small II} region was fully resolved. Second, we do not resolve individual stars (except for Pop.~III stars). A large fraction of high-redshift star formation is expected to occur in dense stellar clusters \citep[e.g.][]{Belokurov2023,Claeyssens2026} where a large fraction of massive stars can be ejected \citep[e.g.][]{Oh2016} into the lower density medium. This process is not modeled in our simulations (although see \citealt{Kimm2014,Ma2015,Andersson2023}), but could provide a way for non-equilibrium effects to become important, even if the stellar birth clouds are not disrupted by feedback. 

Finally, we emphasize that non-equilibrium effects do not necessarily preclude high equivalent width emission lines. First, not every star in the galaxy needs to exist in low-density gas. As can be seen in Figure~\ref{fig:ebg_dens}, some stars remain at very high ISM gas densities close to $10^4\ {\rm cm^{-3}}$. We emphasize that only a fraction of star particles need to be in low density gas for $\beta$ to become bluer. Figure~\ref{fig:muv_beta} shows that H$\beta$ equivalent widths can be as high as 350~\AA\ for EBGs. Second, the EW of strong rest-frame optical lines does not scale linearly with the volume of the nebular gas as one may naively assume. This is because in the rest-frame optical the nebular emission can dominate the continuum \citep[e.g.][]{Byler2017,Katz2024_BJ}. In the models shown in Figure~\ref{fig:beta_time}, at the youngest ages, the nebular emission represents 60\% of the total continuum and is 1.6$\times$ stronger than the stellar emission at 6563~\AA, the rest-frame wavelength of H$\alpha$. If we consider a density of $0.1\ {\rm cm^{-3}}$, where the ionization front only reaches 86\% of the Strömgren radius in 2~Myr, even though the nebular continuum will be suppressed by a factor of 0.63, the equivalent width of H$\alpha$ is only suppressed by 18\%, i.e.\ the EW drops slower than the contribution of the nebular continuum to the observed continuum. Hence our models are not in conflict with the observations of \cite{Saxena2024} who discovered EBGs with similarly extreme equivalent widths. 

Since the low-density gas is key for non-equilibrium effects to impact observed UV slopes, one obvious question is why not just measure the density directly from the emission lines. The primary issue here is that very few emission line ratios are sensitive to the densities that are relevant for non-equilibrium effect to matter. Likewise, the emission may be dominated by the dense gas if the low-density emission is suppressed. For example, if we estimate the gas density of the EBGs using [O~{\small II}] emission, we find densities between $100-500\ {\rm cm^{-3}}$ which is typical of H~{\small II} regions. Having a multi-component nebula, as seen in observations \citep[e.g.][]{Harikane2025} complicates density inferences.

In this work, we have considered all emission within 25\% of the virial radius of the halo. We have checked that this assumption has negligible impacts on our conclusions (i.e.\ by integrating all emission out to the virial radius for a subset of EBGs, see Appendix~\ref{app:aperture}); however, there is a third effect that can lead to extremely blue UV slopes in low $f_{\rm esc}$ galaxies that we have yet to discuss. If the nebular gas is spatially offset from the stars, it may cause an observational bias towards blue $\beta$. This can happen in photometric surveys if the nebular emission is too low surface brightness or for spectroscopic observations if the slit does not cover the entire galaxy. Such effects are more often considered in the context of Ly$\alpha$ \citep[e.g.][]{Blaizot2023,Jiang2024}, although for low-density gas, they may also be important for nebular emission more generally. Both offset nebular emission and high $f_{\rm esc}$, may occur simultaneously as \cite{Choustikov2024a} showed that for simulated galaxies with $f_{\rm esc}>20\%$, the H$\alpha$ and Ly$\alpha$ become more extended than the UV emission (see also \citealt{MR2017}).

In conclusion, we have demonstrated that extremely blue UV slopes do not uniquely imply high escape fractions and that the timescale for an ionization front to reach the Strömgren radius might impact out interpretation of high-redshift galaxy spectra.

\section*{Acknowledgments}
This work made extensive use of the dp265 and dp016 projects on the DiRAC ecosystem. HK is particularly grateful to Christopher Mountford and Alastair Basden for support on DIaL3 and Cosma8, respectively. HK and the {\small MEGATRON} team are especially thankful for the support on Glamdring provided by Jonathan Patterson. The material in this manuscript is based upon work supported by NASA under award No. 80NSSC25K7009. TK is supported by the NRF of Korea (RS-2022-NR070872 and RS-2025-00516961) and partially supported by the Yonsei Fellowship, funded by Lee Youn Jae. FC acknowledges support from a UKRI Frontier Research Guarantee Grant (PI Cullen; grant reference EP/X021025/1).

This work used the DiRAC@Durham facility managed by the Institute for Computational Cosmology on behalf of the STFC DiRAC HPC Facility (\url{www.dirac.ac.uk}). The equipment was funded by BEIS capital funding via STFC capital grants {\small ST/P002293/1}, {\small ST/R002371/1} and {\small ST/S002502/1}, Durham University and STFC operations grant {\small ST/R000832/1}. This work also used the DiRAC Data Intensive service at Leicester, operated by the University of Leicester IT Services, which forms part of the STFC DiRAC HPC Facility. The equipment was funded by BEIS capital funding via STFC capital grants {\small ST/K000373/1} and {\small ST/R002363/1} and STFC DiRAC Operations grant {\small ST/R001014/1}. DiRAC is part of the National e-Infrastructure.

The authors thank Romain Teyssier for both developing and open-sourcing {\small RAMSES}. We thank the developers and maintainers of \textsc{pynbody} (\citealt{Pontzen2013,Pontzen2022}), \textsc{NumPy} (\citealt{vanderWalt2011, Harris2020}), \textsc{SciPy} (\citealt{Virtanen2020}), \textsc{jupyter} (\citealt{RaganKelley2014}), \textsc{matplotlib} (\citealt{Hunter2007}), the Astrophysics Data Service and the arXiv pre-print repository for providing open-source software and services that were used extensively in this work. 

\appendix
\section{Aperture Effects}
\label{app:aperture}
\begin{figure}
  \centering
    \includegraphics[width=0.45\columnwidth]{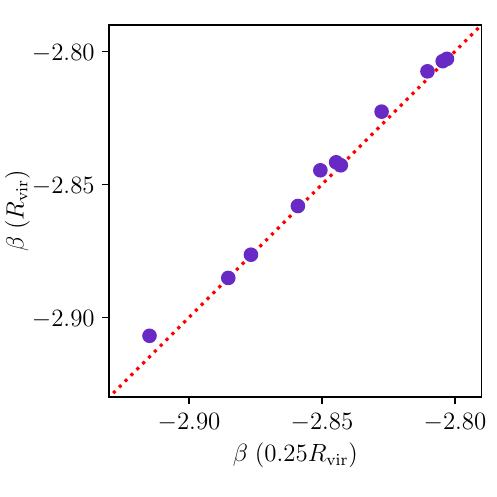}
    \caption{Comparison of UV slope measured within an aperture covering the full virial radius of each halo as a function of the UV slope measured within an aperture of $0.25R_{\rm vir}$. We show the 11 brightest EBGs with $\beta<-2.8$ and M$_{\rm UV}<-18.5$. The dashed red line represents the one-to-one relation. In all cases, the aperture does not impact the $\beta$ measurement indicating that the nebular emission is not offset.}
    \label{fig:beta_beta}
\end{figure}

For our fiducial analysis, we have measured the spectra within $0.25R_{\rm vir}$ and escape fractions at $0.75R_{\rm vir}$. Measuring $f_{\rm esc}$ at $R_{\rm vir}$ would only reinforce our results as the escape fraction will continue to decrease as a function of distance from the center of the galaxy. In contrast, if the nebular emission is offset outside of $0.25R_{\rm vir}$, then our measured UV slopes may be artificially too blue. To demonstrate that this is not the case, we have computed the spectra of the brightest EBGs with $\beta<-2.8$ and M$_{\rm UV}<-18.5$ including all gas cells and star particles within $R_{\rm vir}$. In Figure~\ref{fig:beta_beta}, we compare the UV slopes for the spectra measured within our fiducial aperture and including everything within $R_{\rm vir}$. The two measurements of $\beta$ are nearly indistinguishable, indicating that our chosen aperture does not impact our results.  

\section{Impact of the reduced speed of light approximation}
\label{app:rsla}
To reduce the numerical cost of running the radiative transfer on-the-fly in the simulation. We have artificially reduced the speed of light \citep{Gnedin2001} to $0.01c$. If the ionization fronts propagate faster than our adopted light speed, this will artificially delay the ionization fronts from reaching the Strömgren radius at the correct time \citep[see][]{Rosdahl2013}.

\begin{figure}
  \centering
    \includegraphics[width=0.45\columnwidth]{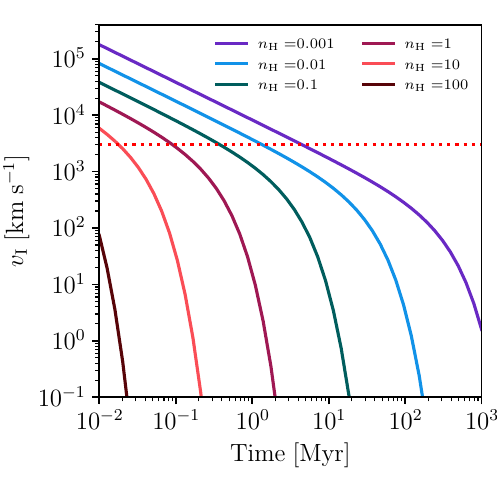}
    \includegraphics[width=0.45\columnwidth]{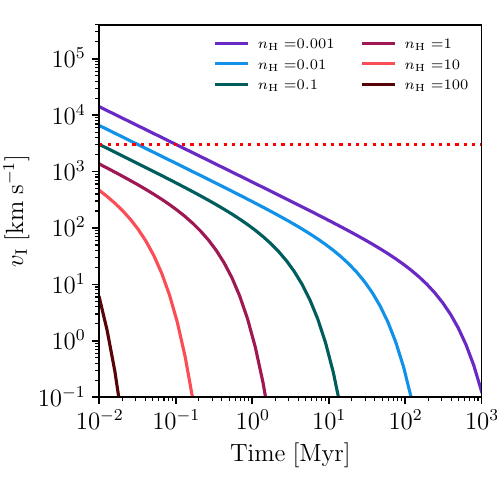}
    \caption{Ionization front velocity as a function of time for star clusters of mass $10^6\ {\rm M_{\odot}}$ (left) or $500\ {\rm M_{\odot}}$ (right) emitting into gas of different densities (colors). The dotted red line shows the value of the speed of light used in the simulation. When the ionization front velocity is moving slower than $c_{\rm sim}$, the detailed numerics do not impact the physics in the simulations. For our star particle mass (right panel), the timescales where the reduced speed of light becomes important are still much smaller than the time it takes for the ionization fronts to reach the Strömgren radius.}
    \label{fig:rsla}
\end{figure}

In Figure~\ref{fig:rsla} we show the ionization front velocity for star clusters of $10^6\ {\rm M_{\odot}}$ (left) or $500\ {\rm M_{\odot}}$ (right) emitting into gas of different densities as a function of time. The more massive the star cluster is, the faster the ionization front moves at a fixed gas density (i.e.\ $v_{\rm I}$ depends on $Q$). The star particles in our simulations are integer multiples of $500\ {\rm M_{\odot}}$ and we find that the reduced speed of light will have negligible impact on our results if the star particles form individually. For very dense star clusters, the reduced speed of light approximation may impact our results for gas densities below 0.1~cm$^{-1}$, which is when the ionization front velocity remains higher than the reduced value for timescales approaching the lifetimes of the most massive stars. However, these timescales where the reduced speed of light becomes important are still much smaller than the time it takes for the ionization fronts to reach the Strömgren radius which is key. Regardless of the numerical details of the simulations, our conclusions are not affected.

\bibliographystyle{mn2e}
\bibliography{example}

\end{document}